\documentclass[pre,aps,showpacs,twocolumn]{revtex4}
\usepackage{amsfonts}
\usepackage{amsmath}
\usepackage{amssymb}
\usepackage{graphicx}
\usepackage{epsfig}
\usepackage{bm}
\begin{document}

\title{A Method to Modify RMT using Short-Time Behavior in Chaotic Systems}

\author{A. Matthew Smith and Lev Kaplan}%
\affiliation{Department of Physics, Tulane University, New Orleans, Louisiana 70118}%


\begin{abstract}
We discuss a modification to Random Matrix Theory eigenstate statistics,
that systematically takes into account the non-universal short-time behavior of chaotic systems.
The method avoids diagonalization of the Hamiltonian, instead requiring only a knowledge of short-time
dynamics for a chaotic system or ensemble of similar systems.
Standard Random Matrix Theory and semiclassical predictions are recovered in
the limits of zero Ehrenfest time and infinite Heisenberg time, respectively.
As examples, we discuss wave function autocorrelations and cross-correlations,
and show how the approach leads to a significant
improvement in accuracy for simple chaotic systems where comparison can be made with brute-force diagonalization.
\end{abstract}

\pacs{05.45.Mt, 03.65.Sq}

\maketitle

The statistical structure of chaotic wave functions has been a key topic of investigation from the early
history of quantum chaos and wave chaos physics, and its study is essential for improved understanding of resonances, transport, and long-time dynamics in non-integrable systems~\cite{mirlin00}. Random Matrix Theory (RMT)~\cite{mehta}, which often serves as an adequate zeroth-order approximation for wave function statistics in the absence of integrability, describes a statistical ensemble of Hamiltonians having no preferred basis. Within RMT, eigenstates are simply random vectors either in the full Hilbert space or in the subspace given by energy and other conservation laws. For a quantum particle in a slowly-varying potential, a wave function then behaves locally like a random superposition of plane waves of fixed wave number, as discussed by Berry~\cite{berry77}.

As a universal theory, RMT specifically excludes system-specific behavior associated with dynamics, boundary conditions, or interactions. Well-recognized deviations from random wave function statistics are associated with boundary effects~\cite{urbina07,biesbdy}, finite system size~\cite{urbina07}, unstable periodic orbits~\cite{scar}, diffusion~\cite{mirlin00}, and two-body random interactions in many-body systems~\cite{tbre,kota}. Much progress has been made in understanding such deviations in various situations of physical interest, for example chaotic wave function correlations in Husimi space associated with classical dynamics~\cite{schanz} and realistic mesoscopic S-matrices arising from a simple diffusive ray picture of wave propagation\cite{weaver}. In particular, semiclassical methods~\cite{srednicki} have proven very successful in quantifying the effects on wave functions of boundaries~\cite{urbina07,biesbdy} and periodic orbit scars~\cite{scar}. However, the limit implied by semiclassical approximations may not always be achievable or relevant in describing actual experiments. For example, an analysis of electron interaction matrix elements in ballistic quantum dots
shows that even for thousands of electrons in the dot, several statistical quantities of interest typically exceed random wave predictions by a factor of 3 or more; for other quantities the random wave model fails even to predict the correct sign~\cite{alhassid} (see also \cite{tomsovic}).

In some situations, e.g.,~\cite{alhassid}, brute force diagonalization of the Hamiltonian may be used to obtain correct statistics for the stationary or long-time behavior, but for very large Hilbert spaces, such as those that arise in many-body situations, diagonalization is likely to be impractical. Even where it ``works'', diagonalization is unlikely to produce much intuition about the relevant physics, and must be repeated for each new Hamiltonian. In fact, individual eigenstates of a chaotic Hamiltonian are highly sensitive to perturbations of the system, particularly for multi-particle systems. The {\it statistics} of such systems are far more robust and remain accurate for small perturbations.

Our goal here is to present a system and basis-independent way of supplementing RMT with short time dynamical information, that eliminates the need for diagonalization of the Hamiltonian, and that provides greatly improved accuracy over RMT and semiclassical methods for finite systems with a finite Ehrenfest time.

To enable direct comparison with RMT, let us consider fully chaotic (ballistic or diffusive) dynamics without symmetry on an $N$-dimensional Hilbert space with eigenstates $|\xi\rangle$. To avoid ambiguities in the definition of $|\xi\rangle$,
we assume a non-degenerate spectrum. Typical quantities of interest, then, are functions of the amplitudes $\langle a|\xi\rangle$ for any physically-motivated basis state $|a\rangle$, which may be a position or momentum state, a Slater determinant, or more generally an eigenstate of some zeroth-order Hamiltonian. With the normalization $\sum_{\xi=1}^N |\langle a|\xi\rangle|^2=1$, the simplest and first non-trivial moment of these amplitudes is given by the local inverse participation number (IPR), which measures the degree of localization at $|a\rangle$:
\begin{equation}
\label{paa}
P^{aa} = N\sum_{\xi=1}^N |\langle a|\xi\rangle|^4 = N \lim_{T \to \infty} \frac{1}{2T}\int_{-T}^{T} dt\,
|\langle a|a(t)\rangle|^2 \,,
\end{equation}
varying from $P^{aa}=1$ in the case of perfect ergodicity to $P^{aa}=N$ for perfect localization. For two arbitrary states we have
\begin{equation}
\label{pab}
P^{ab} \!= \!N\sum_{\xi=1}^N |\langle a|\xi\rangle|^2 |\langle b|\xi\rangle|^2 \!= \! N \lim_{T \to \infty} {1 \over 2T}\!\int_{-T}^{T}\! dt\,
|\langle a|b(t)\rangle|^2 \,.
\end{equation}
Obviously, higher-order moments and in general the entire joint distribution of the eigenstate intensities may be considered (e.g.,~\cite{bootstrap}). We may also relax the requirement that only pure states such as $|a\rangle\langle a|$ act as probes,
and instead measure the structure of chaotic eigenstates using any desired self-adjoint operator $\hat \alpha$~\cite{eckhardt}. Operator probes (of phase space size greater than or smaller than $\hbar$) will, for example, be particularly helpful in the study of hierarchical eigenstates in a mixed chaotic-regular phase space~\cite{hier}. Again, without loss of generality we may adopt the normalization ${\rm Tr} \,\hat \alpha=1$. Eq.~(\ref{pab}) becomes
\begin{equation}
\label{palphabeta}
P^{\alpha\beta} \!=\! N\sum_{\xi=1}^N \langle\xi|\hat \alpha|\xi\rangle  \langle\xi|\hat \beta|\xi\rangle= N \lim_{T \to \infty} {1 \over 2T}  \! \int_{-T}^{T}dt\,
{\rm Tr} \,\hat \alpha \hat\beta(t) \,,
\end{equation}
with the autocorrelation $P^{\alpha\alpha}$ as an obvious special case.

In the semiclassical limit $N \to \infty$, averages of the form (\ref{paa})-(\ref{palphabeta}) may be obtained using short-time dynamics; specifically for discrete-time dynamics we have
\begin{equation}
\label{pabsclimit}
\overline{P^{ab}} \approx \overline{\sum_{-\tau}^{\tau} P^{ab}(t)} \,,
\end{equation}
where
\begin{equation}
P^{ab}(t)=|\langle a|b(t)\rangle|^2+\langle a|a(t)\rangle \langle b(t)|b\rangle \,,
\end{equation}
where $|a\rangle$ and $|b\rangle$ may be any two states (identical, overlapping, or orthogonal) \cite{bootstrap}.
Here and in the following, $\overline{\cdots}$ indicates an ensemble average. If desired, the ensemble may be
selected so that all realizations possess the same short-time dynamics $P^{ab}(t)$, in which case the average
on the right hand side of (\ref{pabsclimit}) is superfluous. The cutoff time $\tau$ 
must be long compared to the ballistic or diffusive Thouless time (so as to include all the non-universal dynamics), and short compared to the Heisenberg time, which scales with $N$. No distinction is made in (\ref{pabsclimit}) between non-universal short-time revivals that indicate deviations from RMT in the eigenstate statistics and the $O(1/N)$ short-time revivals that are present already in the context of RMT. 
As a result, (\ref{pabsclimit}) systematically overestimates corrections to RMT, and violates probability
conservation $\sum_b P^{ab}=1$ given a complete basis $|b\rangle$ for any $\tau >0$, with the violations growing
linearly as $\tau/N$. 

We now notice that the problematic aspects of (\ref{pabsclimit}) for finite system size $N$ can be eliminated by introducing a $\tau$- and $\langle a|b\rangle$-dependent prefactor:
\begin{equation}
\label{pabfactor}
\overline{P^{ab}} \approx C_{N}^{\langle a|b\rangle}(\tau) \overline{\int_{-\tau}^{\tau} dt\, P^{ab}(t)} \,,
\end{equation}
where in particular $C_{N}^{\langle a|b\rangle}(\tau)=N/4\tau$ converges to the exact answer as $\tau \to \infty$. To fix $C_{N}^{\langle a|b\rangle}$, we apply RMT to Eq.~(\ref{pabfactor}) and obtain
\begin{equation}
\label{ratio}
\overline{P^{ab}} \approx \overline {P^{ab}_{\rm RMT}} {\overline{\int_{-\tau}^{\tau} dt\, P^{ab}(t)}
\over \;\;\;\overline{\int_{-\tau}^{\tau} dt\,P^{ab}_{\rm RMT}(t)} \;\;\;}\,.
\end{equation}
Eq.~(\ref{ratio}), and its natural extensions to higher-order moments (e.g., $\overline{(P^{ab})^n}$) and operator expectation values (e.g., $\overline{P^{\alpha\beta}}$) are a key result of this paper. Stationary eigenstate
properties of an quantum chaotic system or ensemble of systems may be fully described by a combination of short-time dynamics for that system or ensemble, in combination with exact results from RMT, without any need
for matrix diagonalization. Reassuringly, Eq.~(\ref{ratio}) yields exact results in three limits of interest: (i) the RMT limit where $P^{ab}(t)=P^{ab}_{\rm RMT}(t)$ and thus $P^{ab}=P^{ab}_{\rm RMT}$, (ii) the semiclassical limit $N/\tau \to \infty$, where we recover
(\ref{pabsclimit}), and (iii) the limit where an infinite amount of dynamical data is available as input, $\tau \to \infty$. More importantly, as we will see in the examples below, Eq.~(\ref{ratio}) and its extensions
provide reliable approximations to exact diagonalization in situations far from any such limit, i.e., for finite-size systems far from universality, and where the only input is short-time dynamics on the scale
of a Lyapunov time. 

Short-time overlaps $P^{ab}(t)$ needed as input to Eq.~(\ref{ratio}) may sometimes be known analytically, as in the case of periodic orbit scars, while in more general situations the short-time dynamics for a given system of interest is easily obtainable numerically, to any desired time scale $\tau$. The RMT factors in (\ref{ratio}) and its generalizations may be treated entirely analytically. For example, for arbitrary $|a\rangle$ and $|b\rangle$ we have standard results in the absence of time reversal symmetry (GUE or CUE)
\begin{equation}
\overline{P^{ab}_{\rm RMT}} = \frac{N}{N+1} \left(1 +|\langle a|b\rangle|^2 \right)\,,
\end{equation}
while for a general self-adjoint operator
$\hat \alpha$ we obtain
\begin{equation}
\label{palphaalpharmt}
\overline{P^{\alpha\alpha}_{\rm RMT}} = \frac{N}{N+1} \left(2 \sum_i A_i^2+\sum_{i \ne j} A_iA_j \right) \,,
\end{equation}
where $A_i$ are the eigenvalues of $\hat \alpha$ ($\sum_i A_i=1$).

Similarly, RMT dynamical overlaps may be expressed exactly using RMT eigenstate statistics and the
RMT spectral form factor, e.g., 
\begin{align}
&\overline{P^{ab}_{\rm RMT}(t)}=\frac{2}{N} \overline{P^{ab}_{\rm RMT}}+
\sum_{\xi \ne \xi'}(\overline{e^{i(E_{\xi'}-E_\xi)t}})_{\rm RMT} \nonumber \\
&\times  \left(\overline{|\langle a|\xi\rangle|^2 |\langle b|\xi'\rangle|^2}+\overline{\langle a|\xi\rangle
\langle \xi|b\rangle\langle \xi'|a\rangle\langle b|\xi'\rangle}\right)_{\rm RMT}
\end{align}
For discrete-time dynamics, described by the CUE ensemble, which will be relevant for the numerical examples below,
we have
\begin{equation}
\overline{P^{ab}_{\rm RMT}(t)}= (1+|\langle a|b\rangle|^2) \times \begin{cases} 1 &  \text{for~}t= 0\\
{1+t/N \over N+1} & \text{for~} 1\le |t|\le N\\
{2 \over N+1} & \text{for~} |t| > N \end{cases} \,,
\end{equation}
and analogous results for self-adjoint operators are obtained by spectral decomposition, as in
(\ref{palphaalpharmt}).

Thus, eigenstate statistics for a chaotic system or ensemble of systems may be unambiguously obtained without
diagonalization, as in
(\ref{ratio}), by combining exact RMT results with easily obtainable short-time dynamical information for the system
or ensemble of interest.

\begin{figure}[h]
{
\psfig{file=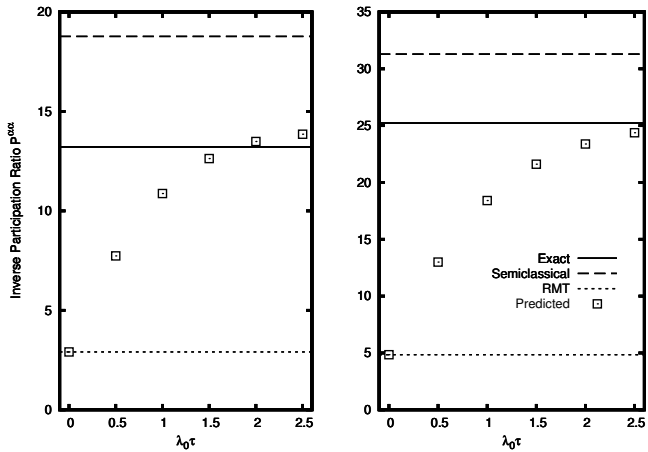,width=0.5\textwidth,angle=0}
}
	 \protect\caption{The inverse participation ratio $P^{\alpha\alpha}$  for a Gaussian distribution centered on
	 a short periodic orbit with instability exponent $\lambda_0=0.5$ is computed by direct diagonalization and
	 compared with  the short-time dynamical prediction given by Eq.~(\ref{ratio}). Here the system size is $N=32$ and the Gaussian distribution has size $s=0.5$ (Left panel) or $s=0.25$ (Right panel). Convergence to the exact result is observed when the dynamical calculation includes information about times $\tau$ up to $2$ or $3$ in units of the local Lyapunov exponent $\lambda_0$.
The RMT value $P^{\alpha\alpha}_{\rm RMT}=(1+s^{-1})N/(N+1)$ and the semiclassical result $P^{\alpha\alpha}_{\rm SC}=(1+s^{-1})(N/(N+1))\sum_{t=-\infty}^{\infty} \text{sech}(\lambda_0 t)$ are shown for comparison.	 
	 All quantities appearing here and in subsequent figures are dimensionless.}
		\label{fig_ipralphat}
\end{figure}	

We now discuss a few illustrative examples, using as our model the paradigmatic example of a quantized periodically kicked Hamiltonian~\cite{kicked}
\begin{equation}
H(q,p,t)=T(p)+V(q)\sum_{n=-\infty}^\infty \delta(t-n)
\end{equation}
on the compact phase space $(q,p) \in [-1/2,1/2)^2$.
The kinetic and potential terms are chosen to produce a fully chaotic map (perturbed cat map~\cite{boasman})
\begin{align}
T(p)&={ m\over 2}p^2+{ K\over 4\pi^2} \cos(2 \pi p)+t(p) \\
V(q)&=-{m \over 2}q^2-{ K\over 4\pi^2} \cos(2 \pi q)+v(q)\,,
\end{align}
where the parameters $m$ and $K$ control the chaoticity of the system: the dynamics is fully chaotic for $m>|K|$ and the
instability exponent of the shortest periodic orbit at $q=p=0$ is $\lambda_0=\cosh^{-1}\left(1+(m-K)^2/2\right) \approx m-K$ for $m-K \ll 1$. To break time reversal and parity symmetries, and also allow for ensemble averaging of the statistics, we have added the functions $t(p)$ and $v(q)$, which are random within a small region near the edges of the phase space ($|p|>1/2-\Delta$ and $|q|>1/2-\Delta$) and zero elsewhere. In the following, we set $\Delta=0.1$, but the results have no significant dependence on $\Delta$.

\begin{figure}[h]
\centerline{\includegraphics[width=0.33\textwidth,angle=270]{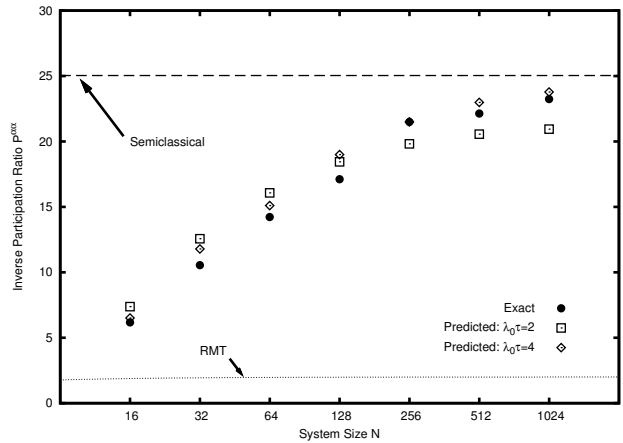}}
	 \protect\caption{
The inverse participation ratio for a pure Gaussian wavepacket ($s=1$) is computed exactly and compared with the dynamical prediction of Eq.~(\ref{ratio}) using dynamical information up to times $\tau=2\lambda_0^{-1}$ and $4\lambda_0^{-1}$, where
$\lambda_0=0.25$ is the local Lyapunov exponent. Results are shown for various values of the system size $N$. The semiclassical and RMT limits are also shown for comparison (see Fig.~\ref{fig_ipralphat} caption).}
		\label{fig_ipralpha}
\end{figure}

We begin by considering the inverse participation ratio $P^{\alpha\alpha}$, where $\hat \alpha$ is the Weyl transform of a Gaussian distribution $\rho(q,p)\sim e^{-q^2/\sigma_q^2-p^2/\sigma_p^2}$ centered on the periodic orbit. We define $s=\sigma_q\sigma_p/\hbar$. Then in the special case $s=1$, $\hat \alpha$ is a projection onto a minimum uncertainty Gaussian wave packet, while more generally $\hat \alpha$ represents a mixed initial state. Typical results are shown in Fig.~\ref{fig_ipralphat}, where the dynamical prediction of Eq.~(\ref{ratio}) for several values of the cutoff time $\tau$ is compared with exact values obtained by brute-force diagonalization. We note that the dynamical prediction begins at the RMT limit for $\tau=0$, as it
must, and quickly converges to the exact stationary answer at $2$ or $3$ Lyapunov times. Fig.~\ref{fig_ipralpha} illustrates the relationship between the exact value of the inverse participation ratio, the dynamical prediction, and the limiting RMT and semiclassical approximations, as the system size $N$ is varied. Here we note significant deviations from the semiclassical answer even when $N$ takes values of $100$ or greater; these deviations are well reproduced in the dynamical calculation.

As another example, we consider wave function intensity correlations $P^{ab}$ for position states $|a\rangle$, $|b\rangle$. Since  ${1 \over N^2}\sum_{a,b=1}^N \overline{ P^{ab}}=1$ is given by wave function normalization when $\tau \to \infty$, we focus on the first interesting moment, the variance
\begin{equation}
\label{v}
W={1 \over N^2} \sum_{a,b}\overline{ (P^{ab})^2}-1\,.
\end{equation}
$W$ is a simple measure of non-uniformity in infinite-time transport~\cite{wqe}, and ranges from $W=0$ for perfect ergodicity to $W=N-1$ for prefect localization. We note also that interchanging the roles of eigenstates and basis states,
$W$ may be equivalently written as the variance of the interaction matrix elements $P^{\xi\xi'}=N\sum_{a=1}^N |\langle a|\xi\rangle|^2 |\langle a|\xi'\rangle|^2$ between eigenstates $|\xi\rangle$ and $|\xi'\rangle$, i.e.,
$W={1 \over N^2} \sum_{\xi,\xi'}\overline{ (P^{\xi\xi'})^2}-1$.
The statistics of such interaction matrix elements in chaotic systems frequently appear in applications ranging
from quantum dot conductance in the Coulomb blockade regime~\cite{alhassid} to controlling directional emission properties in microcavity lasers~\cite{turecistone}.

\begin{figure}[h]
\centerline{\includegraphics[width=0.33\textwidth,angle=270]{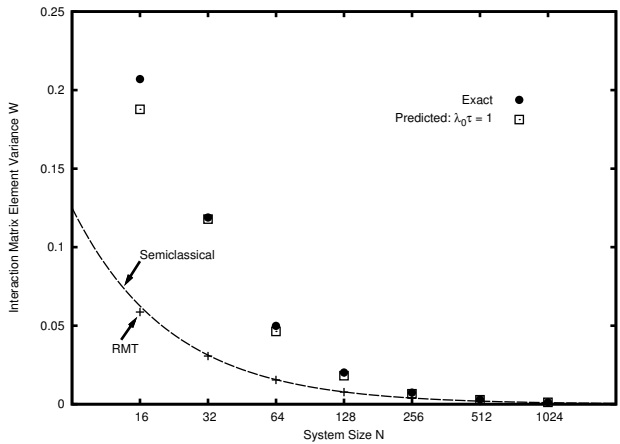}}
	 \protect\caption{
The interaction matrix element variance $W$ is computed exactly (Eq.~(\ref{v})) and compared with the short time prediction (\ref{ratio-2}), with $\tau\lambda_0=1$. Here $\lambda_0=0.125$. The RMT result and the semiclassical limit
$W_{\rm SC}=1/N$ are also shown for comparison. }
		\label{fig_ime}
\end{figure}

We again combine short time dynamics and RMT to calculate the variance of the interaction matrix elements, similarly to Eq.~(\ref{ratio}),
\begin{equation}
\label{ratio-2}
\overline{(P^{ab})^2} \approx \overline {(P^{ab}_{\rm RMT})^2} \; {\;\;\;\overline{\left(\int_{-\tau}^{\tau} dt\, P^{ab}(t)\right)^2}\;\;\;
\over \overline{\left(\int_{-\tau}^{\tau} dt\,P^{ab}_{\rm RMT}(t)\right)^2} }\,.
\end{equation}

We note here that the intensity correlators $P^{ab}$ predicted by Eq.~(\ref{ratio}) are not guaranteed to satisfy the normalization condition $\frac{1}{N^2}\sum_{a,b=1}^N \overline{P^{ab}}=1$, that holds for the exact correlators. This normalization is only guaranteed for (Eq.~(\ref{ratio})) when $\tau \to \infty$. In order to predict $W$ we already need knowledge of the short-time dynamics, $P^{ab}(t)$, for every pair of initial and final states $|a\rangle$, $|b\rangle$; therefore with little added computational effort we may achieve exact normalization and further improve the convergence with $\tau$, simply by rescaling $\overline{(P^{ab})^2} \to \overline{(P^{ab})^2} /(\frac{1}{N^2}\sum_{a',b'=1}^N \overline{P^{a'b'}})^2$.

Fig.~\ref{fig_ime} shows that the semiclassical and RMT predictions are very similar for the system we consider here, and both deviate significantly from the exact results for finite $N$. Our method, including rescaling, converges toward the exact answer very quickly, on the order of the Lyapunov time, even where the RMT prediction is off by a factor of 2 or 3. The accuracy can be improved further by calculating the dynamics for longer times $\tau$.

We have developed a method that improves on RMT eigenstate statistics for chaotic systems by systematically  incorporating short-time dynamics. The method is conceptually appealing, computationally simpler than brute-force diagonalization, and significantly more accurate than RMT or the semiclassical limit for realistic systems. The approach can be easily extended to consider symmetry effects (including time reversal symmetry), mixed phase space~\cite{backer}, and resonance wave function statistics in open systems.

\begin{acknowledgments}
This work was supported in part by the NSF under Grant No.\ PHY-0545390. 
\end{acknowledgments}

\end{document}